\documentclass{mem}
\usepackage{natbib} \usepackage{txfonts} \usepackage{balance}
\usepackage{graphicx}
\usepackage[a4paper]{hyperref}

\begin{document}

\title{
3D spectroscopy of  candidate double-barred lenticular galaxies
}
\author{A.\,V.\, Moiseev }
  \offprints{Alexei Moiseev}

\institute{
Special Astrophysical Observatory, Nizhnij Arkhyz, 369167, Russia
\\
\email{moisav@gmail.com, moisav@sao.ru}
}
\authorrunning{Moiseev}

\titlerunning{3D spectroscopy of  double-barred galaxies}

\abstract{
Significant fraction of barred galaxies hosts secondary  bar-like structures on  optical and NIR images. The circumnuclear  dynamics of  double-barred objects are still not well understood, observational data concerning   kinematics are incomplete and inconsistent.   In order to compare the simulations results with  observations, we have started new spectroscopic  studying  of stellar kinematics in lenticular galaxies from Peter Erwin's catalog of secondary bars. We present  first results concerning their stellar kinematics based on the  observations performed with the integral-field spectrograph MPFS at the Russian 6-m telescope.
}
\maketitle{}

\section{Introduction}

Traditionally, an inner bar-like structure appeared inside a large-scale bar on direct images of barred galaxies call as ``a secondary bar''.  Interest to this problems significantly picked up after the idea by \citet{shlosman89},  that nested-bar systems might  fuel nuclear activity by  driving gas into the nuclear regions of a galaxy. \citet{erwin08} has presented on this workshop an excellent review of secondary bar properties, see also \citet{erwin04}, \citet{mois02} and \citet{mois04}. The main result in study of photometric properties is that ``a secondary bar seems as a small copy of a large-scale counterpart''. However a secondary bar origin, kinematics  and dynamics is not such clear.
Here we briefly outlined some main problems and discrepant judgements about double-barred (hereafter - DB) galaxies. The new observational results concerning S0-galaxies will also be considered.

\section{What is frequency of  such structures in galaxies?}

In 1990s  several small samples, each with about dozen DB candidates, were published.  Based on the literature  \citet{mois01} have listed 71 galaxies where possible secondary bars were suspected from isophotal analysis. Really  a number of objects in this list are not truly DB systems, because various  structural features produce similar distortion  of inner isophotes in barred galaxies (circumnuclear  disks and spirals, triaxial or oblate bulges, etc.). From these reasons \citet{erwin04}  presented a catalog of 50 DB galaxies, based on detailed examinations of previously known as well new found  candidates. A numerous old candidates were rejected, first of all because they have inner disks instead secondary bars. This re-examination has changed significantly some previous estimations of  relative DB fractions. So, \citet{laine02} in their larger statistical study of the secondary bars frequency have found 19 DB systems out of  112 galaxies (69 barred). However the cross-identification with the Erwin's catalogue shows that only six candidates are confirmed as DB\footnote{Also \citet{erwin04} marks 4 objects in \citet{laine02} sample as ``ambiguous'' objects: Mrk1066, NGC\,2339, NGC\,4750 and NGC\,7742. Here the turn of isophotes is caused by spirals or rings, moreover, NGC\,7742 is a classical example of a ringed galaxy  without bar \citep{sil06,muz06}.}. Therefore the fraction of secondary bars in barred galaxies denoted by \citet{laine02} as $28\pm3\%$ should be significantly decrease to $9\pm2\%$. A large percentage $26\pm7\%$ was suggested by \citet{erwin02} in the sample of  S0-Sa bright barred galaxies and all candidates were confirmed in \citet{erwin04}. Their sample is twice smaller than in \citet{laine02} and contains only early-type galaxies. It's possible that fraction of DB is larger in case of early  types; in any case  kinematic arguments   should be also involve into the verification of all DB candidates.

\section{3D spectroscopy of DB: the history}

Since the motions of stars and gas inside the bar region are strongly non-circular, and such objects are non-axisymmetric by definition,  the 2D maps of kinematical parameters seem helpful in the study of DB kinematics. Therefore the  3D (panoramic) spectroscopy  in optical domain or high-resolution  radio  maps are needed. Unfortunately these data are rare and incomplete. For instance, several attempts of studying  molecular gas kinematics in secondary bars were described, but mainly without certain results. So,  \citet{petit02} did not detect any specific kinematic features in the inner region of NGC\,2273, whereas  \citet{erwin02} and MVC04 showed this galaxy harbors an inner disk nested in the large-scale bar. NGC\,5728 \citep{petit02} and NGC\,4303 \citep{Schinnerer02} also haven't   non-circular motions, may be because the size of their inner bars are comparable with width of the telescope beam. The secondary bar in NGC\,2782 \citep{hunt08} does not exert on the observed gas velocity field, however the numerical simulations of morphology required  to include  such structures in its model. The similar  situation is also presented in the case of NGC\,4579 \citep{garcia08}  and for the ionized gas kinematics in NGC\,3359 \citep{rozas08}. However   clear examples of kinematically decoupled secondary bars are found  for the molecular gas. Namely,  only the central regions of  Maffei\,2 \citep{meier08} and Our Galaxy \citep{rodrigues08}  exhibit  the remarkable features  on $P-V$ (or $l-v$) diagrams which are reproduced in the secondary bar models.

The first stellar velocity fields of DB candidates were mapped with OASIS integral-field spectrograph: NGC\,3504 and NGC\,5850 \citep{eric00}, NGC\,470 \citep{eric02}, NGC\,2273 \citep{eric01}. However, only short notes in some proceedings were published, without detailed analysis. Moreover,
 \citet{erwin04} and MVC04 show that last two objects have  inner disks
inside the large bars. The SAURON sample of early-type galaxies contains only two candidates from the Erwin's catalogue: NGC\,4314 which does not show any velocity deviations in the center \citep{sauronVII} and NGC\,1068 \citep{eric06} where kinematically decoupled inner bar was detected, however the outer bar (oval) is still under discussion.

The larger collection of kinematical maps for 13 DB candidates based on 3D spectroscopic observations were published by us (MVC04). And we detected the dynamical decoupling of circumnuclear ($r\sim1$ kpc) regions in all observed galaxies, but one (NGC\,5566). Various complex structures (coplanar or
polar mini disks, mini-spirals, etc.) were revealed. A large fraction of
peculiar structures is not surprising, because different structural
features can produce   twists of the inner isophotes. However, we have not found any  kinematic features expected for the secondary bars. Also we have observed only  6 ``confirmed candidates'' from the Erwin's catalogue, some of them have ambiguous morphology. For instance, his ``strong candidate'' NGC\,3368 has a polar disk nested in the small bar and  spiral arms erroneously interpreted early  as a large-scale bar \citep{silmois03}.

Therefore  new 3D spectroscopic observations are required together with a careful selection of the best candidates.

\section{Why we prefer S0?}

Now we choose for  the observations  only lenticular galaxies from the \citet{erwin04} list, because  their  morphology and relatively small gas and dust contributions simplify a problem of inner bars detections. For instance  $\sim50\%$ of objects  in \citet{erwin04} are S0 and S0/a. Also the  detailed   self-consistent collisionless simulations of secondary bars originated from rotating pseudobulges in  early-type galaxies were recently considered \citep{debshen07,shen07}. The stellar density and kinematics maps   constructed by \citet{shen07}  allow us to make a comparison between integral-field data and their numerical simulations.

\section{New MPFS observations}

The observations were carried out in December, 2007 - March, 2008 at the Russian 6-m telescope with the integral-field spectrograph MPFS \citep{afan01}. The field of view (FOV) was $16''\times16''$ with scale $1''/$spaxel (or $0.5''$ in the case of  drizling for  NGC\,6654) and mean seeing $1.3''-1.6''$. We took stellar velocities and velocity dispersion maps ($\sigma$-maps)  in 6 S0-S0/a galaxies with  sizes of secondary bar $2''<L_2<6''$ that provides an enough   sampling for the FOV. We also add in our analysis 2 SB0  galaxies NGC\,2950 and NGC\,3945 observed early with the same device (MVC04).

\begin{figure*}[t!]
\centerline{\includegraphics[width=6.5 cm]{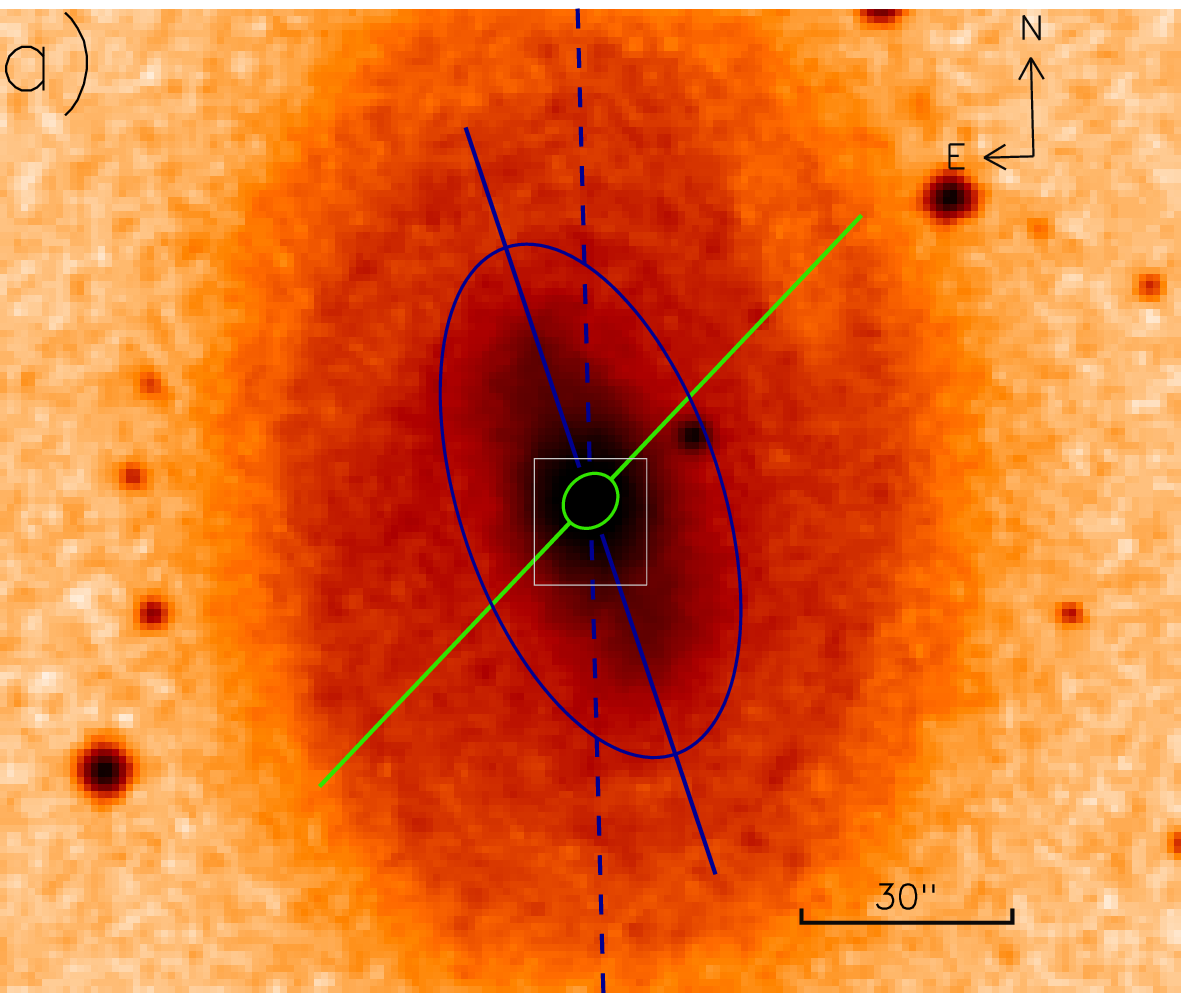}}
\centerline{\includegraphics[width=6 cm]{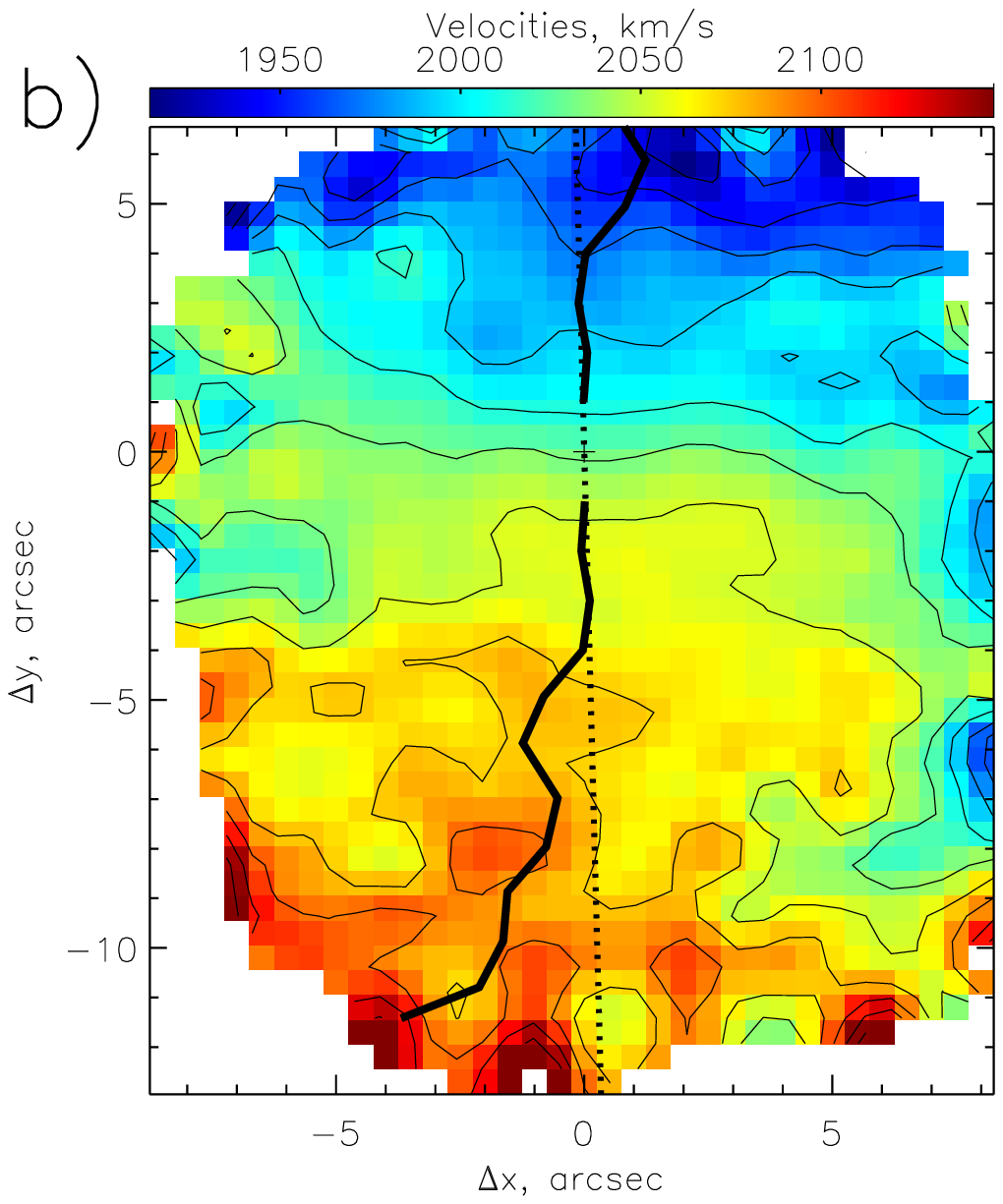}
\includegraphics[width=6 cm]{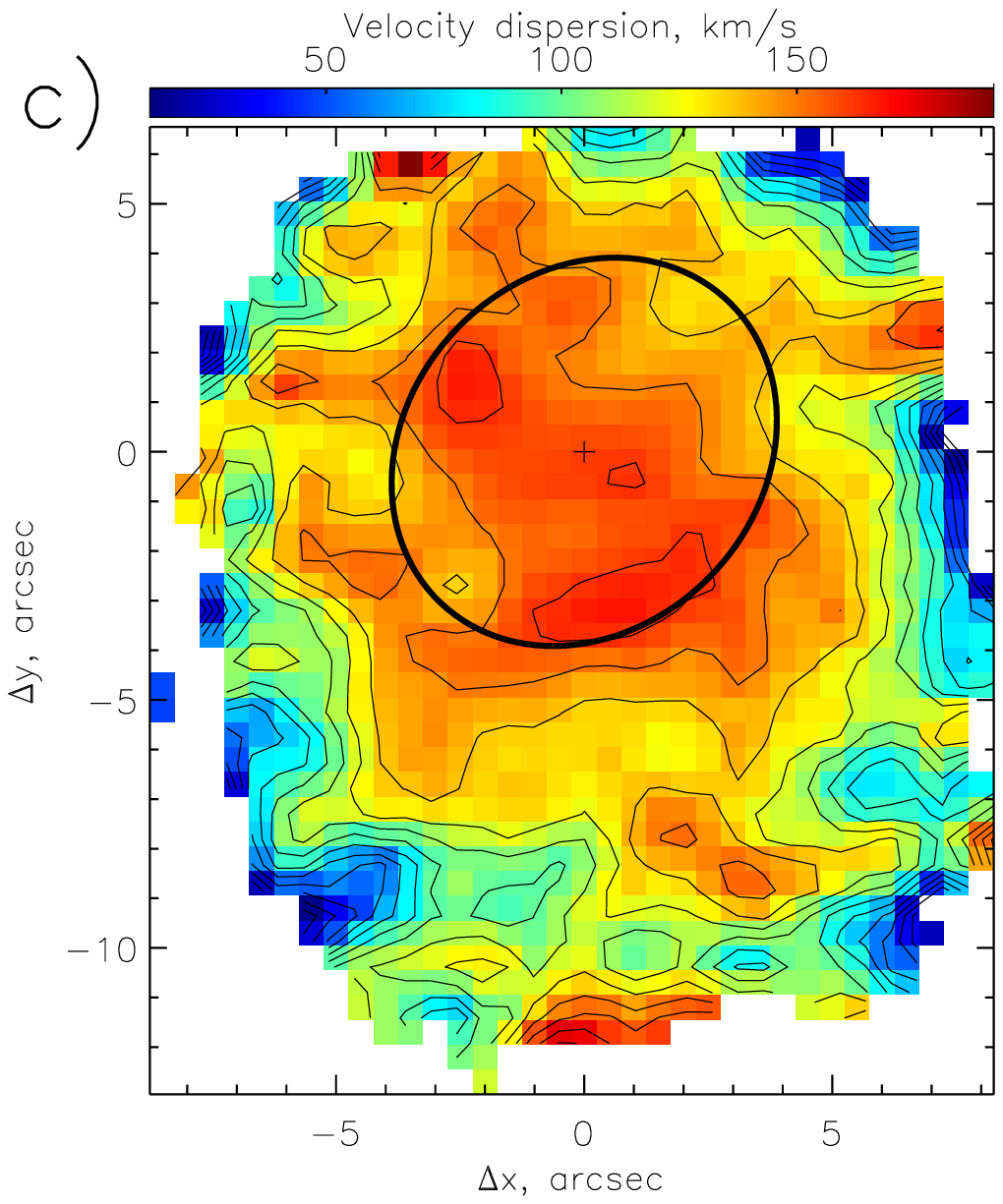}}

\caption{\footnotesize S0/a galaxy NGC\,6654. \textbf{a)} DSS2  image, ellipses and straight  lines mark the orientation of both bars in agreement with their length, PA and ellipticity  from \citet{erwin04}. The dashed line shows  line-of-nodes position. The white rectangle labels the combination of two MPFS FOVs. \textbf{b)} Velocity field of the stars. Dotted line is the disk line-of-nodes, solid line shows the position of kinematic major axis. \textbf{c)} Map of the stellar velocity dispersion. The ellipse shows the orientation and size of the inner bar.
}
\label{fig1}
\end{figure*}

\subsection{Velocity fields}

In the case of single-barred galaxy the non-circular motions  in non-axisymmetric potential distort the line-of-sight velocity field. The well-known effect is a twist of the kinematical major axis in opposite direction with respect  to the bar position angle. The N-body simulation by \citet{shen07} shows that in secondary bar the  deviation of position angle ($\Delta PA$) twice smaller than in the case of a single bar (see their Fig.7). Of  course, this value depends on disk and bar orientation. Nevertheless, the typical value of the distortion $\Delta PA=10-15^\circ$ significantly larger than uncertainty of $PA$ estimations derived from a MPFS velocity field, so the isovelocity twist should be detected.

Fig.~\ref{fig1} shows the MPFS maps for  NGC\,6654. The kinematical axis derived from the velocity  field coincides with disk position angle at $r<4.5''$. In other words, a pure circular regular rotation is observed  on the radii corresponded to the secondary bar. The   twist of the kinematical axis ($\Delta PA=7-20^\circ$) in agreement with the orientation  of the large scale bar is presented at $r>5''$, just outside the photometric borders of the ``inner bar''. In all galaxies in our sample we observe the  same behaviour of kinematic axis: \textit{an absence  of non-circular motions on the scale of photometrical secondary bar}.

The regular disk-like circumnuclear kinematics cannot be explained by the beam effect of angular resolution, because the photometrical length of the secondary  bars are significantly larger than seeing during the observations. Moreover,  with the MPFS we already detected nuclear mini-bars (without large-scale counterparts) in some galaxies, for instance NGC\,3368 \citep{silmois03} or NGC\,3786 \citep{mois02}.  Second possibility is  that  rotation of the bright bulge are observed along the line-of-sight and suppressed the possible contribution of the secondary bar in the total velocity field. Though \citet{shen07} models also include a fast rotating bulge, they  predicted the detectable twist of kinematic axis. It will be interesting to compare our observations with simulations of galaxies with the same  bulge/disk ratio and relative orientations of the bars as in the  observed galaxies.

\subsection{Velocity dispersion and $\sigma$-hollows}

Simple models of single-barred galaxies have shown \citep{mil79,vd97} that the central ellipsoidal peak in the $\sigma$-maps must be aligned with the  bar major axis. In real galaxies the distribution of velocity dispersion is more complex, along line-of-sight we see the contribution of different stellar population in several dynamical components. Usually (in $\sim50\%$ of objects) observations revealed  $\sigma$-drop \citep{emsellem01}, lopsided and amorphous  structures instead symmetrical  peaks. Despite this problem, if an elliptical peak exists,  the $PA$ of the elliptical peak has  a strong
correlation with the $PA$ of the large-scale bar major axis \citep{mois02}. In our new sample 5 out of 8  lenticular  galaxies have central elliptical peaks on the $\sigma$-maps. And we confirm the result of our previous work:
\begin{itemize}
\item Peak in the velocity dispersion maps is aligned with the  direction of outer large-scale bar.
\item  There is no  correlation  between $\sigma$-peaks and secondary bar major axes.
\end{itemize}

Therefore, the large-scale bar drives the stellar motions even into the regions
where the photometric inner bars are observed.

Recently, \citet{lorenzo08} published new SAURON data for stellar kinematics in 4 candidates DB. At the ends of photometric inner bars they found  local minima of the velocity dispersion ($\sigma$-hollows) and supposed a connection of this features with a dynamically cold  inner bar where high-ordered stellar motions are present. We happened to observe 2 galaxies from  their sample and confirm the $\sigma$-hollows in NGC\,2859. Despite the  smaller FOV of the  MPFS,  we found similar features in NGC\,2950 and NGC\,6654 (see Fig.~\ref{fig1}). However the interpretation offered by  \citet{lorenzo08} seems debatable. Why `more ordered motions' don't exert on velocity fields and appear only in $\sigma$ distributions? It's interesting that the $\sigma$-hollows are mainly observed only in the galaxies with a significant (larger than $60^\circ$) angle between the inner and outer bars. Therefore this feature  \textit{can be connect with the minor axes of outer bar, never with the major axis of inner bar.} Unfortunately, it's a problem to check this idea, because we have very scanty  information  about velocity dispersion behaviour in the outer regions of a large-scale bar. Most of   integral-field maps for barred galaxies   usually have insufficient FOV or relatively low signal in the outer regions. One exception is SAURON map for NGC\,3489 \citep{sauronIII} where $\sigma$-hollows are clearly visible. It seems reasonable that this pattern  is a result of disk velocities redistribution under acts of a bar. Some recent simulations of stellar kinematics in barred galaxies can support this idea \citep{chakra04}, see also Fig.~10 in \citet{vd97}. If the $\sigma$-hollows were produced by large bar,  then the secondary  bar don't distort a circumnuclear stellar kinematics in agreement with the conclusion based on analysis of the velocity fields. In any case, new detailed    simulations of velocity dispersion distribution in barred galaxies are needed.

\section{Conclusion}

We have presented our preliminary results based on analysis of velocity and velocity dispersion fields of stellar component in the circumnuclear regions of 8 candidate double-barred early-type galaxies:

\begin{itemize}
\item The twist of kinematical axis in the region of inner bar is absent. As minimum it's significantly  smaller than prediction of modern N-body simulations. This can be connected with large contribution of fast rotating bulge. Nevertheless we have detected primary bar signatures just outside of small bar.

    \item The primary bar drives the stellar velocity dispersion in the
circumnuclear region.

\end{itemize}

Therefore, in contrast with photometric data, where a secondary bar appears as a small copy of a large scale one, the  motions of stars in secondary bars   differ essentially from single-bar kinematics. Also it's possible that  the fraction of kinematically  decoupled bars is still overestimated when we use only photometric (morphological) criteria. Probably, that real kinematically decoupled secondary bar is a rare phenomenon  connected with other peculiarities, like bars counter-rotation offered by \citet{mac06}.

\begin{acknowledgements}
I wish to thank  Olga Sil'chenko, Witold Maciejewski and Dalia Chakrabarty  for helpful discussions and Aleksandrina Smirnova for some comments.   This work was supported by the Russian Foundation for Basic Research (project  06-02-16825) and by the grant of President of Russian Federation (MK1310.2007.2).
\end{acknowledgements}

\bibliographystyle{aa}

\begin{thebibliography}{}

\bibitem[{Afanasiev, Dodonov, \& Moiseev(2001)}]{afan01}
Afanasiev V.\ L., Dodonov S.\ N.~\& Moiseev A.\ V.  2001, in
 ``Stellar dynamics: from classic to modern'' (eds. Osipkov L.P., Nikiforov
 I.I.)  Saint Petersburg,  103

\bibitem[Chakrabarty(2004)]{chakra04} 	
Chakrabarty, D. 2004, \mnras, 352, 427

\bibitem[Debattista \& Shen(2007)]{debshen07}
Debattista, V. P.  \& Shen, J. 2007, \apj, 654, L127

\bibitem[de Lorenzo-C{\'a}ceres et al.(2008)]{lorenzo08}
de Lorenzo-C{\'a}ceres, A., Falc{\'o}n-Barroso, J. Vazdekis, A., Mart{\'{\i}}nez-Valpuesta, I. 2008, \apj, 684, L83

\bibitem[Emsellem(2002)]{eric02} Emsellem, E. 2002, ASP Conf. Ser., 282, 145, astro-ph/0202522

\bibitem[Emsellem \& Friedli(2000)]{eric00}  Emsellem, E. \& Friedli, D. 2000, ASP Conf. Ser., 197, 51, astro-ph/9910260

\bibitem[Emsellem et al.(2001a)]{emsellem01} Emsellem, E., Greusard, D., Combes, F., et al. 2001a, \aap, 368, 52

\bibitem[Emsellem et al.(2001b)]{eric01} Emsellem, E., Greusard, D., Friedli, D., Combes, F. 2001b, ASP Conf. Ser., 230, 235

\bibitem[Emsellem et al.(2004)]{sauronIII} Emsellem, E., Cappellari, M., Peletier, R. F. 2004, \mnras, 352, 721

\bibitem[Emsellem et al.(2006)]{eric06} Emsellem, E., Fathi, K., Wozniak, H. et al. 2006, 365, 367


\bibitem[Erwin(2004)]{erwin04} Erwin, P.\ 2004, \aap, 415, 941

\bibitem[Erwin(2008)]{erwin08} Erwin, P.\ 2008, this volume

\bibitem[Erwin \& Sparke(2002)]{erwin02} Erwin, P.~\&   Sparke, L.\ S. 2002, \aj, 124, 65

\bibitem[Falc\'{o}n-Barroso et al.(2006)]{sauronVII}Falc\'{o}n-Barroso, J., Bacon, R., Bureau, M. 2006, \mnras, 369, 529

\bibitem[Garc\'{i}a-Burillo et al.(2008)]{garcia08} Garc\'{i}a-Burillo, S. Fern\'{a}ndez-Garc\'{\i}a, S.,  Combes, F. et al. 2008, \aap, sumbitted, arXiv:0810.4892 [astro-ph]

\bibitem[Hunt et al.(2008)]{hunt08} Hunt, L. K.,  Combes, F., Garc\'{i}a-Burillo, S. et al. 2008, \aap, 482, 133

\bibitem[Laine et al.(2002)]{laine02} Laine, S., Shlosman, I., Knapen, J.H., Peletier, R.F. 2002, \apj, 567, 97

\bibitem[Maciejewski(2006)]{mac06}Maciejewski, W. 2006, MNRAS, 371, 451

\bibitem[Mazzuca et al.(2006)]{muz06}
Mazzuca, L.\ M.,  Sarzi, M.,  Knapen, J.\ H., Veilleux, S., Swaters, R. 2006, \apj, 649, L79

\bibitem[Meier, Turner   \& Hurt (2008)]{meier08} Meier D. S., Turner, J. L. \& Hurt, R. L. 2008,  ApJ, 675, 281

\bibitem[Miller   \& Smith (1979)]{mil79} Miller, R.H.  \& Smith, B.F.  1979, \apj, 227, 785

\bibitem[Moiseev(2001)]{mois01} Moiseev, A.V., 2001, Bull. Spec. Astrophys. Obs., 51, 140, astro-ph/0111220

\bibitem[Moiseev(2002)]{mois02} Moiseev, A. V.  2002, Astronomy Letters, 28, 755, astro-ph/0211105

\bibitem[Moiseev, Vald{\'e}s, \& Chavushyan (2004, hereafter MVC04)]{mois04}
Moiseev, A.\ V., Vald{\'e}s, J.\ R. \& Chavushyan, V.\ H.\  2004, \aap, 421, 433 (MVC04)

\bibitem[Petitpas \& Wilson(2002)]{petit02} Petitpas, G.\ R.~\&  Wilson, C.\ D. 2002, \apj, 575, 814

\bibitem[Rodriguez-Fernandez  \& Combes(2008)]{rodrigues08} Rodriguez-Fernandez N. J. \& Combes F. 2008, \aap, 489, 115

\bibitem[Rozas(2008)]{rozas08} Rozas, M. 2008, RMxAA, 44, 71

\bibitem[Schinnerer et al.(2002)]{Schinnerer02} Schinnerer, E., Maciejewski, W., Scoville, N., Moustakas, L.A. 2002, \apj, 575, 826

\bibitem[Shen \& Debattista(2007)]{shen07}
Shen, J. \& Debattista, V. P.   2007, \apj, submitted (arXiv:0711.0966)

\bibitem[Shlosman, Frank, \& Begeman(1989)]{shlosman89}
Shlosman, I., Frank, J.~\&  Begeman, M.C. 1989, Nature, 338, 45

\bibitem[Sil'chenko \&  Moiseev(2006)]{sil06}
Sil'chenko, O.\ K.~\&  Moiseev, A.\ V. 2006, \aj, 131, 1336

\bibitem[Sil'chenko et al.(2003)]{silmois03}
Sil'chenko, O.\ K.,  Moiseev, A.\ V., Afanasiev V. L. et al. 2003, \apj, 131, 1336

\bibitem[Vauterin  \& Dejonghe(1997)]{vd97} Vauterin, P., \& Dejonghe, H. 1997, \mnras, 286, 812
\end{thebibliography}

\end{document}